# Momentum transfer in a standing optical vortex

V. G. Shvedov

*Nonlinear Physics Centre, Research School of Physical Sciences and Engineering,*
*the Australian National University, Canberra, ACT 0200 Australia*
e-mail: vlad-shvedov@yandex.ru

*and*

*Physical Department, V.I. Vernandsky Taurida National University,*
*Vernadsky av. 4, Simferopol, 95007, Crimea, Ukraine*

## Abstract

A field superposition of singular beams incident on, and then reflected from a mirror has been investigated. It was demonstrated that the standing optical wave, which contains a vortex, possesses an orbital angle momentum where the energy flux circulates only in the azimuth direction of the beam. We show in this paper that the standing light wave containing the optical vortex transfers angular momentum to a substance located in the field of the vortex without moving the substance in the azimuth or radial directions. This property of the standing vortex present an opportunity to form the three-dimensional optical traps, gasdynamic and hydrodynamic vortices, in a localised volume by a direct transfer of the orbital angular momentum from the optical vortex.

## 1. Introduction

Singular optical beams have been known since the laser radiation wave function was presented as a steady-state solution to the Schrödinger wave equation for a harmonic oscillator [1]. Furthermore, the behaviour of the singular beams in free space has been studied from the time mathematical modelling of the beams was developed [2]. Recently, singular beams have been considered as a special case of the wave equation's solutions with research into these solutions being the core of a new branch of optics now known as 'singular optics' [3,4]. A major step in understanding the nature of singular beams was made by Allen et. al., who considered local energy transfer trajectories in optical vortices [5]. This provided a means to consider the orbital angular momentum as an intrinsic feature of the optical vortex. However, the orbital angular momentum was always considered for a travelling wave and the issue of the orbital angular momentum in a standing wave was still an open question. The intention in this paper is to reveal the structure of the orbital angular momentum in a standing wave formed by the optical vortex reflected from a mirror surface.

## 2. Linearly polarised mode of the standing vortex

Let us approach first the propagation of the optical vortex in the free space. The solution of the scalar steady-state paraxial wave equation may be written as:

$$f = \frac{r^{|l|}}{\rho^{|l|}(\sigma)^{|l|+1}} \exp(il\phi) \exp\left(\frac{-r^2}{\rho^2\sigma}\right) \exp(ikz), \tag{1}$$

where $\sigma = 1 + i\dfrac{z}{z_0}$, $z_0 = \dfrac{k\rho^2}{2}$ is the relay length, $\rho$ is the beam waist radius, $l$ is the topological charge of the vortex, and $k$ is the wavenumber. By separating the phase and the amplitude, Eq.(1) can be presented as:

$$f = \left(\frac{z_0}{\sqrt{z^2 + z_0^2}}\right)^{|l|+1} \frac{r^{|l|}}{\rho^{|l|}} \exp\left(\frac{-r^2 z_0 k}{2(z^2 + z_0^2)}\right) \exp\left(il\phi - i\frac{r^2 zk}{2(z^2 + z_0^2)} - i(|l|+1)\arctan\frac{z}{z_0} + ikz\right).$$

The expression in the second exponent provides a way to determine the surface of the beam wave front using the following relation:

$$\Theta = -\frac{r^2 k}{2R} - (|l|+1)\arctan\frac{z}{z_0} + l\phi + kz = const; \tag{2}$$

where $R$ is the radius of curvature of the wave front [6]:

$$R = z\left(1 + \frac{z_0^2}{z^2}\right). \tag{3}$$

The absolute values for the electric field intensity in the paraxial linearly polarised beam are related to the scalar wave function, Eq.(1), in the following way:

$$\begin{cases} |\mathbf{E}| = Af \\ |\mathbf{H}| = c\varepsilon_0 Af \end{cases} \tag{4}$$

where $A$ is the amplitude of the electric field, $\varepsilon_0$ is the dielectric constant, and $c$ is the speed of light in vacuum.

Let us consider the reflectivity from a perfect mirror surface, of an arbitrarily polarised wave with electric and magnetic field as presented in Eq.(4). We place a mirror with radius of curvature $R$ (as in Eq.(3)), so that the axis of symmetry of the mirror is coincident with the $z$-axis of the beam, and the centre of curvature is on the axis at a distance $z'$ from the mirror. The mirror surface is in accordance with the wavefront of Eq.(2). It can be shown from Eq.(3) that the reflectivity of the beam from the mirror, and the centre of curvature in the point $z = z'$, is equivalent to the reflectivity from a flat mirror placed in the plane of the beam waist at $z = 0$ within the accuracy of the beam divergence. The scale of the beam divergence is given by:

$$r^{/} = \frac{r}{\sqrt{1 + \frac{z^2}{z_0^2}}}.$$

The electromagnetic field calculations are performed in the vicinity of the maximum intensity of the incident beam. The Fresnel equations can be applied for the calculations of the reflected wave in this area [7] and the coordinate system should be chosen so that the origin is in the vertex of the mirror surface $z^{/} = 0$. We should also take into account that the radial projection of the normal component of the wavefront (Eq.(2)) is equal to zero at the point of incidence on the mirror, with radius of curvature found by Eq.(3):

$$\nabla(\Theta) = \frac{l}{kr} \mathbf{e}_\tau + \mathbf{e}_n. \tag{5}$$

Then, using local coordinates of the mirror surface, the electric field of the incident wave $\mathbf{E}^i$ reads:

$$\mathbf{E}_\tau^i = b_n f(r^{/}, \phi, 0) \exp\left[-i\left(\omega t - kr^{/} \sin(q_i) + kz\cos(q_i)\right)\right]$$

$$\mathbf{E}_\phi^i = -b_p f(r^{/}, 0, 0)\cos(q_i) \exp\left[-i\left(\omega t - kr^{/} \sin(q_i) + kz\cos(q_i)\right)\right] \tag{6}$$

$$\mathbf{E}_z^i = -b_p f(r^{/}, 0, 0)\sin(q_i) \exp\left[-i\left(\omega t - kr^{/} \sin(q_i) + kz\cos(q_i)\right)\right]$$

where $q_i = q_i(r)$ is the angle of incidence at the point $i$, and $b_n$ and $b_p$ are the normal (perpendicular) and the parallel components of the electric field vector relative to the plane of incidence. In the general case, $b_n$ and $b_p$ are coordinate dependent and thus determine the wave polarisation state at the point of incidence:

$$b_n = b_n(r^{/}, \phi)$$
$$b_p = b_p(r^{/}, \phi)^.$$

Obviously, the relation $b_n^2 + b_p^2 = const$ holds at any point of the wavefront. Taking into account Eq.(4), the constant can be defined as:

$$b_n^2 + b_p^2 = A^2. \tag{7}$$

Now suppose that the mirror reflectance is equal to unity [7], so that

$$g_p = b_p;$$
$$g_n = -b_n; \tag{8}$$

here $g_n$ and $g_p$ are correspondingly the normal and the parallel projections of the electric field of the incidence wave onto the plane of incidence. The components of the electric field of the reflected wave read:

$$\mathbf{E}_r^r = g_n f(r', \phi, 0) \exp\left[-i\left(\omega t - kr' \sin(q_i) - kz \cos(q_i)\right)\right]$$

$$\mathbf{E}_\phi^r = g_p f(r', \phi, 0) \cos(q_i) \exp\left[-i\left(\omega t - kr' \sin(q_i) - kz \cos(q_i)\right)\right] \qquad (9)$$

$$\mathbf{E}_z^r = -g_p f(r', \phi, 0) \sin(q_i) \exp\left[-i\left(\omega t - kr' \sin(q_i) - kz \cos(q_i)\right)\right]$$

The magnetic field of the incident and reflected waves is orthogonal to the corresponding electric field:

$$\frac{H_r}{b_p} = -c\varepsilon_0 \frac{E_r}{b_n}; \qquad \frac{H_\phi}{b_n} = c\varepsilon_0 \frac{E_\phi}{b_p}; \qquad \frac{H_z}{b_n} = c\varepsilon_0 \frac{E_z}{b_p}.$$

Finally, the resulted field can be found from Eqs.(6, 9) and taking Eq.(8) into account:

$$\mathbf{E}_r = -2b_n f(r', \phi, 0) \sin(kz \cos(q_i)) \exp\left[-i\left(\omega t - kr' \sin(q_i) - \frac{\pi}{2}\right)\right]$$

$$\mathbf{E}_\phi = 2b_p f(r', \phi, 0) \cos(q_i) \sin(kz \cos(q_i)) \exp\left[-i\left(\omega t - kr' \sin(q_i) - \frac{\pi}{2}\right)\right] \qquad (10a)$$

$$\mathbf{E}_z = -2b_p f(r', \phi, 0) \sin(q_i) \cos(kz \cos(q_i)) \exp\left[-i\left(\omega t - kr' \sin(q_i)\right)\right]$$

$$\mathbf{H}_r = -2c\varepsilon_0 b_p f(r', \phi, 0) \cos(kz \cos(q_i)) \exp\left[-i\left(\omega t - kr' \sin(q_i)\right)\right]$$

$$\mathbf{H}_\phi = -2c\varepsilon_0 b_n f(r', \phi, 0) \cos(q_i) \cos(kz \cos(q_i)) \exp\left[-i\left(\omega t - kr' \sin(q_i)\right)\right] \qquad (10b)$$

$$\mathbf{H}_z = 2c\varepsilon_0 b_n f(r', \phi, 0) \sin(q_i) \sin(kz \cos(q_i)) \exp\left[-i\left(\omega t - kr' \sin(q_i)\right) - \frac{\pi}{2}\right]$$

Eqs.(10a,b) define all the components of the electromagnetic field of the superimposed incident and reflected waves of the optical vortex.

## 3. Energy flow in the standing vortex wave

The Poynting vector is defined as:

$$P_\phi = -\frac{1}{2}\left[(E_r H_z^* - E_z H_r^*) + (E_r^* H_z - E_z^* H_r)\right]$$

$$P_r = \frac{1}{2}\left[(E_\phi H_z^* - E_z H_\phi^*) + (E_\phi^* H_z - E_z^* H_\phi)\right] \qquad (11)$$

$$P_z = \frac{1}{2}\left[(E_r H_\phi^* - E_\phi H_r^*) + (E_r^* H_\phi - E_\phi^* H_r)\right]$$

thus, using the field relations from Eqs.(10), the Poynting vector components can be expresses as follows:

$$P_r = 4c\varepsilon_0 b_n b_p \left| f(r^/, z^/ = 0) \right|^2 \sin(q_i)\cos(q_i)\left[ \cos^2(kz\cos(q_i)) - \sin^2(kz\cos(q_i)) \right]$$

$$P_\phi = 4c\varepsilon_0 \left| f(r^/, z^/ = 0) \right|^2 \sin(q_i)\left[ b_p^2 \cos^2(kz\cos(q_i)) + b_n^2 \sin^2(kz\cos(q_i)) \right]$$

$$P_z = 0$$

(12)

Averaging expressions in Eq.(12) over space coordinates, and taking into account the relations between $b_p$ and $b_n$ in Eq.(7), we arrive to the following:

$$\rangle P_r \langle = 4c\varepsilon_0 b_n b_p \left| f(r^/, z^/ = 0) \right|^2 \sin(q_i)\left[ \frac{1}{2} - \frac{1}{2} \right] = 0$$

$$\rangle P_\phi \langle = 4c\varepsilon_0 \left| f(r^/, z^/=0) \right|^2 \tan(q_i)\left[ \frac{1}{2} b_p^2 + \frac{1}{2} b_n^2 \right] = 2c\varepsilon_0 A^2 \left| f(r^/, z^/=0) \right|^2 \tan(q_i)$$

(13)

We rewrite the wavefunction modulus in cylindrical coordinates with the origin in the beam waist taking into account the translational invariant property of the field $\left| f(r^/, 0) \right|^2$ :

$$\left| f(r^/, 0) \right|^2 = \left| f(r, z) \right|^2;$$

where $\left| f(r, z) \right|$ is presented by Eq.(1). In the paraxial beam case, we derive from the Eq.(5) the maximum intensity of the field

$$\tan(q_i) = \frac{l}{kr}.$$

(14)

Finally, the Poynting vector reads:

$$\langle \mathbf{P} \rangle = \left[ 0, \ \frac{2l}{kr} c\varepsilon_0 A^2 \left| f(r, z) \right|^2 \mathbf{e}_\phi, \ 0 \right].$$

(15)

Hence the optical vortex reflected from a mirror is not a complete standing wave, but a travelling wave along the tangential component of the beam. The tangential component of the standing wave is two times larger than that of the travelling wave with the same topological charge.

## 4. Momentum in the standing vortex wave mode

### 4.1. Normalising the wave energy

Let us determine the coefficient $A$, in the expression for the Poynting vector:

$$P_\phi = c\varepsilon_0 A^2 \frac{2l \left| f \right|^2}{kr};$$

(16)

for this we calculate the energy of a travelling wave crossing the beam per unit time:

$$N = \varepsilon_0 c \iint_\infty |\mathrm{E}|^2 \, dS = \pi \varepsilon_0 c A_l^2 \frac{|l|! \rho^2}{2^{|l|+1}} = const \, . \qquad (17)$$

Taking into account the fact that the energy of radiation in *cw* beam is constant and in case of non-absorbing media is independent of the beam charge, the *A* coefficient reads:

$$A_l^2 = \frac{2^{|l|+1} N}{|l|! \pi \varepsilon_0 c \rho^2} \qquad (18)$$

and thus the final expression of the Poynting vector reads:

$$\mathrm{P}_\phi = \frac{N}{\pi \rho^2} \frac{|f|^2}{kr} \frac{2^{|l|+2}}{|l-1|!} \cdot \qquad (19)$$

### 4.2. Angular momentum of the standing optical vortex

The field momentum density [7] is determined by the Poynting vector in the following way:

$$\mathbf{G} = \frac{\mathbf{P}}{c^2} \, .$$

The field momentum [7] reads:

$$\mathbf{L} = \iiint \mathbf{G} \, dV = \frac{1}{c^2} \iiint \mathbf{P} \, dV \, .$$

In a similar manner, the angular momentum density reads:

$$\mathbf{m} = \frac{1}{c^2} \mathbf{r} \times \mathbf{P} \, , \qquad (20)$$

and the total field of the angular momentum in volume V is:

$$\mathbf{M} = \iiint \mathbf{m} \, dV = \frac{1}{c^2} \iiint \mathbf{r} \times \mathbf{P} \, dV \, . \qquad (21)$$

Integrating Eq.(21) over the volume and taking Eqs.(16,18) into account we obtain the following expression for the z-component of the angular momentum:

$$\mathrm{M}_z = \frac{2\pi h}{c^2} \int r \mathrm{P}_\phi r dr = 2l A_l^2 \frac{2\pi \varepsilon_0 h}{ck} \int |f|^2 r dr = 2l A_l^2 \frac{\pi \varepsilon_0 h}{ck} \frac{|l|! \rho^{2|l|+2}}{2^{|l|+1}} = 2lh \frac{N}{kc^2} \qquad (22)$$

The derived solution

$$\mathrm{M}_z = 2lh \frac{N}{kc^2} \qquad (23)$$

represents the angular momentum of the standing wave in the volume of a beam path with length *h* and unlimited beam radius.

## 5. Angular momentum transfer to a substance

Let us insert an absorbing substance in front of the mirror. The wave transfers the following momentum over the time $t$:

$$M_z = 2lh\frac{N}{kc^2}C_k = 2lct\frac{N}{kc^2}C_k = 2l\frac{N}{kc}tC_k = 2l\frac{N}{\omega}C_k t \; ; \tag{24}$$

here $C_k$ is the absorption coefficient. It can be found from the following condition:

$$C(z)_k = \frac{N - N_{exit}(h)}{N}\; ; \tag{25}$$

where $N - N_{exit}(h)$ is the energy of the absorbed radiation in a layer of length $2h$, during the time period $t$. The following condition should be fulfilled for the wave to be a standing wave:

$$1 >> C_k(z) \geq 0. \tag{26}$$

However, near the focal area of the incident and reflected vortices this condition becomes non-mandatory. Instead, equalisation of the intensities of both waves in this area may be performed using a ring resonator with a single roundtrip (a resonator without feedback) (Fig.2). The relation Eq.(22) then takes the following form:

$$M_z = \left((-1)^{(m+1)} + 1\right)l\frac{N}{\omega}c_k z t \; ; \tag{27}$$

where $c_k$ is the specific energy absorption coefficient determined by absorption in the media over a unit path length in a unit of time, and $m$ is the number of the reflective surfaces. As is evident from Eq.(27), there should be an odd number of reflections to form a standing wave vortex.

For the purpose of this study, namely, for calculating the angular momentum transfer to a substance, there is no need to consider the whole volume of the substance. We have restricted ourselves to the case of media with zero viscosity, and to the area where the main part of the momentum transfer is in the vicinity of the maximum intensity.

### 5.1. Particles trapping in the maximum intensity in the layer of length h, radius $r_m$ and width $\Delta r$

The momentum transfer in this case is as follows:

$$M_z = \frac{-2\pi h}{c^2}\int_{r_m - \frac{\Delta r}{2}}^{r_m + \frac{\Delta r}{2}} r P_\phi r dr\big|_{\Delta r \to 0} = 2\frac{2^{|l|+2}}{|l-1|!}\frac{N_m h}{k\rho^2 c^2}|f_m|^2 r_m \Delta r \; ; \tag{28}$$

here $r^2_m = \dfrac{l(z^2 + z_0^2)}{kz_0}$. We assume that the intensity does not change significantly in a thin layer

near the maximum:

$$\left| f_m \right|^2 = \left( \frac{z_0^2}{z^2 + z_0^2} \right)^{l+1} \frac{r_m^{2l}}{\rho^{2l}} \exp\left( \frac{-r_m^2 z_0 k}{(z^2 + z_0^2)} \right) = \left( \frac{l}{2} \right)^l \left( \frac{z_0^2}{z^2 + z_0^2} \right) \exp\left( -|l| \right) ;$$

thus we derive the following expressions for the momentum transfer:

$$\mathrm{M}_z = 2l \frac{2^{|l|+2}}{|l|!} \frac{N_m h}{k\rho c^2} \left( \frac{l}{2} \right)^{l+\frac{1}{2}} \exp\left( -|l| \right) \Delta r \approx 2l \frac{N_m h}{k\rho c^2} \Delta r' . \tag{29}$$

Consequently, taking the expansion of the beam with propagation in the media as:

$$\Delta r = \Delta r' \sqrt{1 + \frac{z^2}{z_0^2}} ,$$

Eq.(29) provides a means for calculation of the rotation velocity of small colloidal particles suspended in a media with low viscosity. To do this the angular momentum has to be calculated for all the particles:

$$\mathrm{M}_z = I_{zz} \omega_z = \varpi_z \int \sum_l x_l^2 \rho_c dV . \tag{30}$$

### 5.2. Angular momentum calculations of particles in the field

For calculating the momentum of a single particle in a field, we take:

$$\left| f_m \right|^2 r_m = \rho \left( \frac{l}{2} \right)^{l+\frac{1}{2}} \left( \frac{z_0}{\sqrt{z^2 + z_0^2}} \right) \exp\left( -|l| \right) ;$$

Then Eq.(29) can be presented as:

$$\mathrm{M}_z \approx 2l \frac{N_m h}{k\rho c^2} \left( \frac{z_0}{\sqrt{z^2 + z_0^2}} \right) \Delta r$$

where $\mathrm{M}_z$ is the angular momentum of the field in the layer of thickness $\Delta r$.

Let us assume that $\Delta r$ is the radius of a particle trapped in the maximum intensity of the beam, so that there are $m$ particles in the maximum intensity area:

$$m = \frac{2\pi r_m}{2\Delta r} = \frac{\pi r_m}{\Delta r} .$$

It can be assumed that the particles are of spherical shape. In this case, the field angular momentum in the space where one particle is trapped in the standing vortex is as follows:

$$\mathrm{M}_{1z} = \frac{\mathrm{M}_z}{m} = \frac{\mathrm{M}_z \Delta r}{\pi r_m} ;$$

$$\mathrm{M}_{1z} \approx 2l \frac{N_m}{k\rho c^2} \left( \frac{z_0}{\sqrt{z^2 + z_0^2}} \right) \frac{\Delta r^3}{\pi r_m} ;$$

$$\text{M}_{1z} = 2\sqrt{2l}\,\frac{N_m}{k\rho^2 c^2}\left(\frac{z_0^2}{z^2+z_0^2}\right)\frac{\Delta r^3}{\pi} = \sqrt{2l}\,\frac{N_m}{c^2}\left(\frac{z_0}{z^2+z_0^2}\right)\frac{\Delta r^3}{\pi}.$$

In the vicinity of the beam waist, the momentum is:

$$\text{M}_{1z} = \sqrt{2l}\,\frac{N}{z_0 c^2}\frac{\Delta r^3}{\pi} = 2\frac{\sqrt{2l}}{\pi}\frac{\Delta r^3}{k\rho^2 c^2}N. \tag{31}$$

The value of the angular momentum in the space occupied by the particle is rather small. However, if the particle absorbed all the field energy in that volume, than this presents the angular momentum absorbed within the time interval $\Delta t = \dfrac{2\Delta r}{c}$:

$$\text{M}_{1ell} = 2\frac{\sqrt{2l}}{\pi}\frac{\Delta r^3}{k\rho^2 c^2}N. \tag{32}$$

Under this momentum a particle of radius $\Delta r$ located at a distance $r_m$ rotates in the field with the frequency $\omega_p$:

$$\omega_p = \frac{3\sqrt{2}}{\pi^2}\frac{N}{\rho_p\sqrt{l}}\frac{1}{kc^2\rho^4}; \tag{33}$$

where $\rho_p$ is the particle density. The angular momentum attained by the particle within the time $t$ is:

$$\text{M}_{part} = \frac{\text{M}_{1ell}}{\Delta t}t; \tag{34}$$

The net result is that the angular momentum for the particle in a standing optical vortex is:

$$\text{M}_{part} = \sqrt{\frac{l}{2}}\left(\frac{z_0}{z^2+z_0^2}\right)\frac{\Delta r^2}{\pi c}N_m t. \tag{35}$$

### 5.3. Angular momentum transfer in continuum media

The field intensity on the beam axis is as follows:

$$I_{zz} = \int r^2 \rho_c dV = \rho_c \int r^2 dV = \frac{1}{4}2\pi\rho_c h(r_2^4 - r_1^4).$$

Taking into account $r_2^2 - r_1^2 = \Delta r^2$ and that the layer with maximum intensity is relatively thin, we obtain the following formulas:

$$r_2^4 - r_1^4 = (r_2^2 - r_1^2)(r_2^2 + r_1^2) = 2r_m^2\Delta r^2;$$

$$I_{zz} = \pi\rho_c h r_m^2\left(1+\frac{z^2}{z_0^2}\right)\Delta r'^2;$$

$$\mathrm{M}_z = 2l\frac{N_m t}{k\rho c}c_k h\Delta r' = I_{zz}\omega_c = \omega_c\pi\rho_c hr_m^2\left(1+\frac{z^2}{z_0^2}\right)\Delta r'^2; \qquad (36)$$

$$\omega_c\pi\rho_c hr_m^2\left(1+\frac{z^2}{z_0^2}\right)\Delta r'^2 = 2l\frac{N_m t}{k\rho c}c_k h\Delta r';$$

$$r^2{}_m = \frac{l(z^2+z_0^2)}{kz_0};$$

$$\omega_c\pi\rho_c h\frac{l}{kz_0^3}(z^2+z_0^2)^2\Delta r'^2 = 2l\frac{N_m t}{k\rho c}c_k h\Delta r';$$

$$\omega_c\pi\rho_c\frac{1}{z_0^3}(z^2+z_0^2)^2\Delta r' = 2\frac{N_m t}{\rho c}c_k;$$

$$\omega_c\pi\rho_c\left(1+\frac{4z^2}{k^2\rho^4}\right)^2\Delta r' = 4\frac{N_m t}{k\rho^3 c}c_k.$$

The assumption that the substance layer is located in the half-width (FWHM) of the beam implies:

$$\langle\omega_c\rangle\pi\rho_c\left(1+\frac{4z^2}{k^2\rho^4}\right)^2 = 2\frac{N_m t}{\omega\rho^4}c_k;$$

$$\langle\omega_c\rangle = 2c_k\frac{N_m t}{\pi\rho_c\left(1+\frac{4z^2}{k^2\rho^4}\right)^2\omega\rho^4}.$$

Taking the density of the substance to be 1 kg/m$^3$, the angular velocity of one revolution per second is gained in the time interval of 1 s.

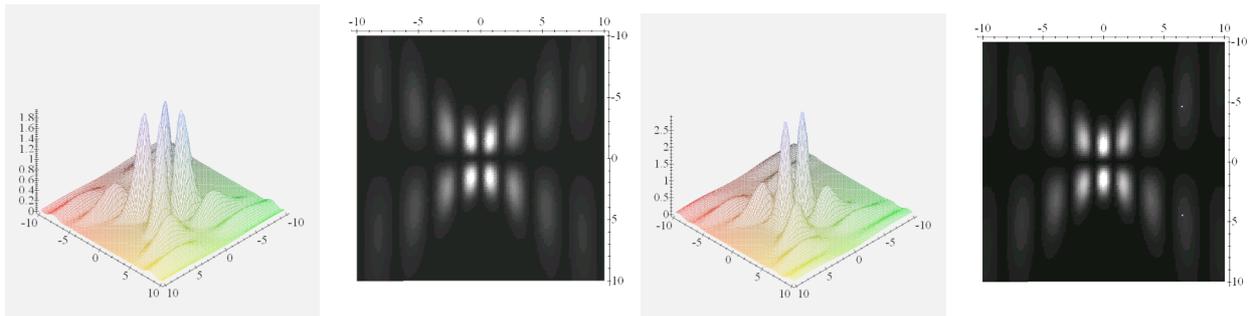

Fig. 1 shows the field intensity profiles: a) E; b) H.

## 6. Conclusions

We have shown theoretically that a reflected optical vortex forms a standing optical wave in both the radial and the longitudinal directions, but is still rotating along the tangential direction. It was demonstrated, that such a standing optical vortex could transfer the doubled angular momentum of the electromagnetic wave to the matter inside in the area of vortex localisation. This momentum transfer from the standing vortex to homogeneous media with low absorption, and to discrete particles, was considered. The transfer of angular orbital momentum to a single particle of a size less than half of the beam size was demonstrated. This study provides the basis on which the three-dimensional optical traps with the gasdynamic or hydrodynamic vortices could be created in a local volume by direct transfer of the angular orbital momentum of the electromagnetic wave to the trapped substance.

## 7. Acknowledgement

## References

1. Kogelnik H., Li T., 1966, Laser beams and resonators, *Appl. Opt.*, **5**, 1550-1567.
2. Oron R., Davidson N., Friesem A. and Hasman E., 2001, Transverse mode shaping and selection in laser resonators, E. Wolf, Progress in Optics, **42,** 325-387.
3. Nye J.F., Berry M.V., 1974, Dislocations in wave trains, *Proc. R. Soc. Lond. A.*, **336**, 165-190.
4. Soskin M., Vasnetsov M., 2001, *Singular optics*, in *Progress in Optics*, E. Wolf, ed., **42**, 219-276.
5. Allen L., Padgett M.J., and Babiker M., 1999, The orbital angular momentum of light, *Progress in Optics*, **39**, 291-372.
6. Yariv A., *Quantum Electronics*, 3rd edition (Wiley, 1989).
7. Born M., Wolf E., *Principles of optics*, 6rd edition (Pergamon Press, 1980).